# Relaxation models for Single Helical reversed field pinch plasmas at low aspect ratio

R. Paccagnella[1,2], S. Masamune[3], A. Sanpei[3]

[1]Consorzio RFX, Corso Stati Uniti 4, 35127 Padova, Italy
[2] Istituto Gas Ionizzati del CNR
[3]Kyoto Institute of Technology, Kyoto, Japan

**Abstract:**

A recent study [1] about the scaling with the aspect ratio (i.e. the ratio between the major and minor radius of the torus) of the dominant mode in Single Helical (SH) Reverse Field Pinch (RFP) plasmas has shown, at intermediate aspect ratio, that the dominant toroidal mode number in the helical states can be interpreted as the result of a relaxation process.

In this work, the theoretical model is compared and validated with the experimental data obtained in the low aspect ratio RELAX device [2].

**Introduction:**

The existence of Single Helical (SH) dominated RFP plasmas, i.e. states with a dominant toroidal mode number, n, and with poloidal mode number m=1, have been observed, since relatively long time both, theoretically and experimentally [3-10]. Note that, since in experiments they are "polluted" by sub-dominant harmonics they are often named "Quasi-Single-Helicity" (QSH) states. Numerical single fluid viscous-resistive magneto-hydro-dynamics (MHD) studies also indicate the existence of local minima in magnetic energy [3] as a possible physical explanation for the appearance of such states.

The famous and elegant Taylor's relaxation theory [11,12] predicts the existence of non axi-symmetric helical states but, as discussed in details for example recently in [13], it fails in finding the observed n numbers. The main obstacle is due to the fact that the Taylor's theory outcome is a flat (constant) parallel current density, while the experimentally observed modes are mostly triggered by the current gradient.

It is well known that the presence of an ideal shell surrounding the plasma is crucial to preserve global invariants (like total helicity, magnetic energy, magnetic flux etc.) [11,12] .

Recently it has also been shown that a toroidal conducting shell is able to respond most efficiently just to the modes observed in RFPs SH states at the different aspect ratios [13]. Therefore for these modes the shell acts almost exactly as an ideal shell.

Also, in recent years, advanced control techniques are applied to RFP plasmas [14,15] helping in achieving the ideal boundary condition of vanishing radial magnetic field at the shell radius and therefore improving the conservation of global quantities.

These observations and findings suggest therefore to reconsider the possibility of explaining the observed helical states as a result of a relaxation process, although slightly different from the Taylor's one. A theory [16-18], proposed a few years after Taylor, and appropriate for cases in which a single mode becomes dominant, has been recently (and successfully) compared with experimental results obtained for intermediate aspect ratios (R/a= 4 - 5 ) [1].

In [1] beside the SH helicity relaxation (SHR) theory (as we call hereafter the model developed in [16]) a two region (TR) model generalizing the Taylor's solution in presence of current sheets was discussed, showing that a cooperative use of the SHR and TR could predict/match the experimental observations regarding the modes that become dominant at different plasma conditions
(i.e. in particular the degree of reversal of the edge toroidal magnetic field ).

In this paper we apply these models to the case of a low aspect ratio RFP, i.e. the RELAX experiment [2]. Moreover, in a recently published paper [19] a method was described to construct the helical symmetric perturbation satisfying the helical Grad-Shafranov equilibrium. The method is applied here to the RELAX case and compared with some magnetic fluctuations data.

The paper is organized as follows: in section I we list some of the important experimental findings regarding the SH states in RFPs and in RELAX; in section II the solutions obtained by the SHR and TR models are described and are compared with the RELAX data; in section III the SHR solution is used to initialize an helical Grad-Shafranov solver in order to obtain more information about the dominant helical symmetric perturbation structure. Finally a discussion and conclusions are given.

*I.    Single Helical Reverse Field characteristics*

Reverse Field Pinches operate in a wide range of aspect ratios (A=R/a, where R and a are the major and the minor torus radii respectively), from 2 to almost 8. At all these different A's single helical states are observed with different toroidal periodicities (n numbers) and with poloidal mode number m=1, as shown in Fig.1.

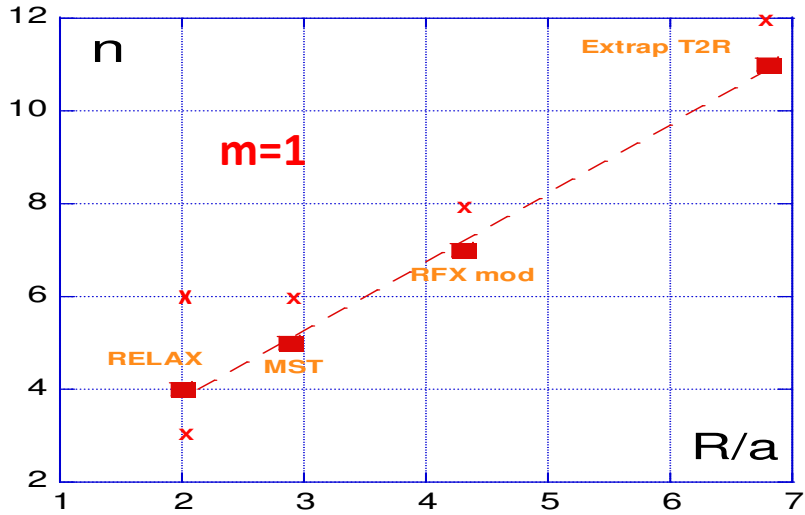

**Fig.1** Dominant n numbers for the SH states at different aspect ratios.

The squares in Fig. 1 refer to dominant modes at shallow reversal while the crosses indicate the dominant mode at deep reversal (see below), apart the cross for the n=3 mode at A=2 which is observed for slightly non-reversed discharges.

Two dimensionless parameters can be defined to characterize different equilibria that, as we shall see in a moment, are also very important in determining the SH characteristics . The reversal of the toroidal field in RFP's is described by the F parameter, i.e. the ratio between the toroidal field measured at the wall and the average toroidal field over the plasma cross section:

$$F = \frac{B_z(a)}{<B_z>} \quad with \ <B_z> = \frac{1}{\pi a^2} \int_0^a B_z \, r \, dr$$

where the average is done in cylindrical geometry and the toroidal field is assumed constant along the boundary. These approximations are reasonable for a low q (safety factor) device like the RFP, where toroidal corrections are small and especially if average quantities are considered, as in this case. Therefore the expression shallow reversal means F near zero (-0.3< F < 0.) , while deep reversal refers to higher negative values of F ( -1 < F < -0.3).

Another important parameter defining the RFP states is the so called pinch parameter Θ , which is:

$$\Theta = \frac{B_\theta(a)}{<B_z>}$$

where $B_\theta(a)$ is the poloidal magnetic field at the wall.

In particular, in RELAX, the edge poloidal and toroidal magnetic fields are measured from top and bottom diagnostic ports at 10-12 toroidal locations, and their averaged values define the edge poloidal field $B_\theta(a)$ and edge toroidal field $B_z(a)$, respectively. The cross-sectional averaged toroidal field $<B_z>$ is obtained from the average of toroidal flux loop signals at 16 equally spaced toroidal locations. We may note that there remains some uncertainty in defining the reversed- or non-reversed discharge arising from toroidal non-uniformity of the edge magnetic fields. In what follows, we have not considered the effect of toroidal asymmetry, which is mainly due to the two insulated poloidal gaps and to the diagnostic ports, in calculating the F-Θ values.

The F and Θ parameters where first introduced in Taylor's theory [11,12] and the F-Θ curves are able to describe in a synthetic and simple way the RFP relaxed states. The experimental F-Θ points are shown in Fig.2 for the RELAX device and for a large number of discharges that include both cases with the emergence or not of the SH states.

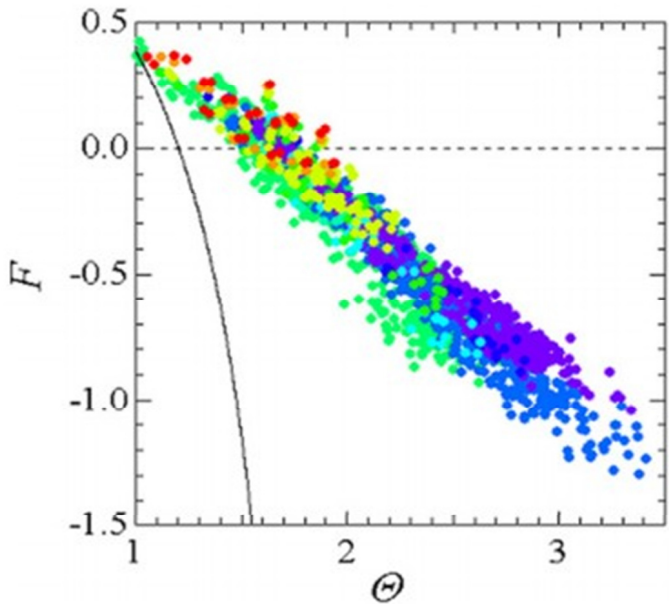

**Fig.2** F vs. Θ relaxed states in RELAX (the plain line is the Taylor's curve) (from Ref.24]).

It can be noted that there is a large amount of data at deep F (and high Θ ) and that the deviation from the Taylor's model prediction is quite strong in this region. It should be also noted (although not shown here) that the deviation is even larger in RELAX than in other RFP's at higher aspect ratios. It can also be seen that the points are well aligned although, for each F value, a moderate spread of the order of around 20% in Θ is present.

In several devices, at different aspect ratios, it has been observed that the dominant mode in SH states changes and in particular it increases (in modulus) with F going from shallow to deep reversal [20,21,22].

Typical measured time traces of the helical structures in RELAX for two different F (and Θ) values are shown in Fig.3. Similar trends are also observed, for different n's, at higher aspect ratios [21,22], where also the dominant modes obtained for the shallower reversal (or even non-reversed) cases persist for longer time intervals with respect to the deep reversed ones.

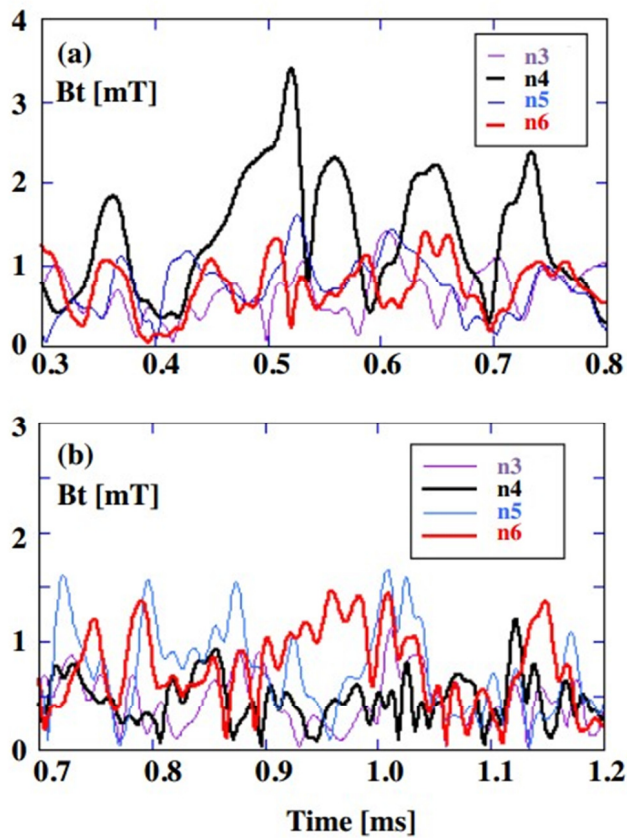

**Fig.3** Toroidal field harmonics vs. time for shallow-reversed (F=-0.1) case (a) and for a deep-reversed case (F=-0.5) case (b).

Generally the helical structures are not long-lived in RELAX, which is a relatively small machine having a fast time scale, the most robust, in this sense, as seen in Fig.3, is the n=4 mode.

Fig.4 (a) shows the dominant n numbers at different F in RELAX employing a bit of statistical analysis.

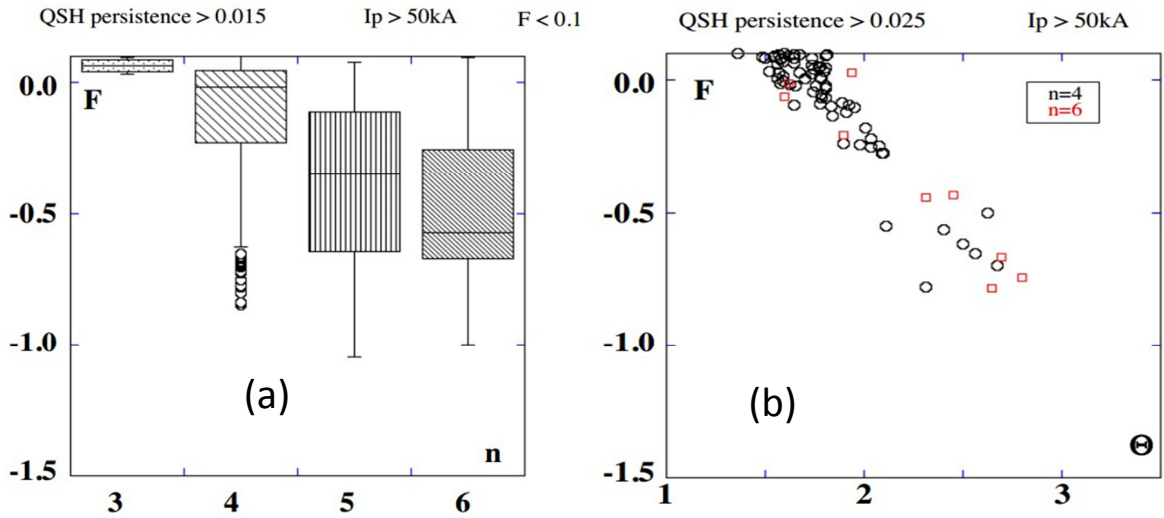

**Fig.4** (a) Dominant toroidal mode number n at different F values; (b) dominant n's on the F-Θ plane; (see text for a more detailed explanation).

The plot is constructed collecting the shots where the appearance of SH state (at various n's) had a persistence higher than 1.5% and the plasma current higher than 50 kA. Here the persistence is defined as the duration of a single event of the SH phase in comparison with the flat-topped current phase. The F-Θ are calculated as the average values during the fla-top.

Each box encloses 50% of the data with the median value of the variable displayed as a line. The top and bottom of the box mark the limits of ± 25% of the variable population. The lines extending from the top and bottom of each box mark the minimum and maximum values within the data set that fall within an acceptable range. Any value outside of this range, called an outlier, is displayed as an individual point.

As shown in Fig. 4(a) at shallow reversal (or even slightly non-reversed cases) the dominant modes are n=3, 4 while at deep reversal n=5,6 could be dominant with slightly similar probabilities.

It can be seen that the data points scatter above F=0, some of which may be attributable to the effect of toroidal asymmetry of the quantities as mentioned previously. In comparing with the theory for the RFP configuration, we will concentrate mainly our attention upon the data points located in the F<0 region. The n=3 becomes dominant only when F is slightly positive [23], maybe suggesting a different nature for this mode (see the discussion at the end of the paper).

The dominant modes plotted in the F-Θ plane are shown in Fig.4(b) where a persistence (as defined above) of about 3% is set as a threshold. It can be seen that in this case only n=4 and marginally n=6 modes persist. Assuming a threshold between 5 and 10% makes only the n=4 to survive the selection.

The last important element is about the parallel current profile that can be deduced from experimental data, by equilibrium reconstruction. As we mention in the Introduction it is expected that the parallel current density defined as:

$$\lambda = \mu_o \frac{\mathbf{J} \cdot \mathbf{B}}{B^2}$$

where $\mu_o$ is the vacuum permeability and **J** and **B** are the current density and the magnetic field, which is constant in the Taylor's case, will vary radially.

Two typical profiles for $\lambda$ are given in Fig.5 as deduced by the equilibrium reconstruction (with the code RelaxFit) in the RELAX device at shallow and deep F values.

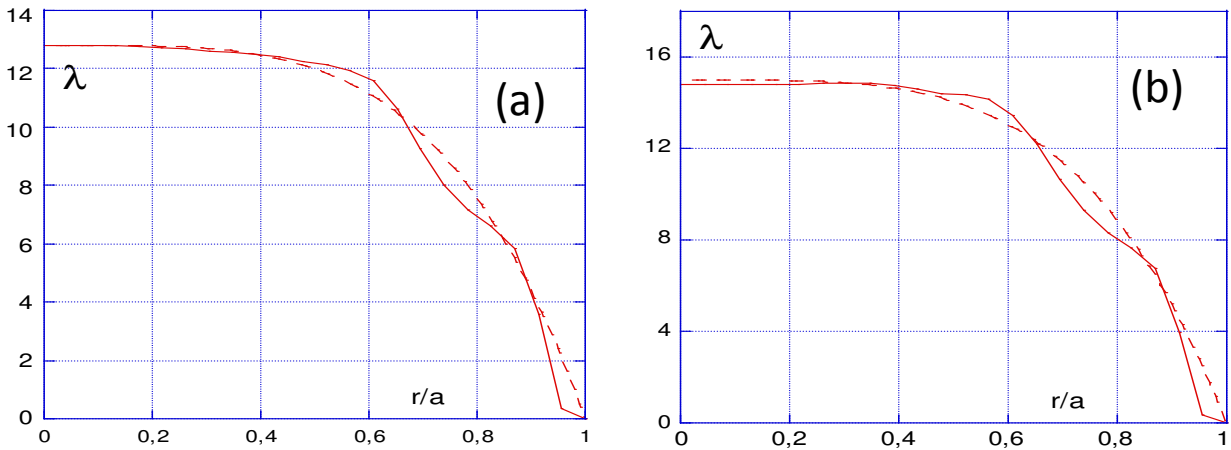

**Fig.5** $\lambda$ profile (in m$^{-1}$) vs. normalized radius for shallow F (a) and deep F (b) values in RELAX. Dashed line represents a fit with $\lambda = \lambda(0) \left[1 - (\frac{r}{a})^4\right]$.

It is seen that the shape of the profile is the same for the two F values (the same function fits both cases) while the on axis value slightly increases with F.

To summarize:

SH helical states are observed in RFPs at all aspect ratios. The dominant mode number n depends on the aspect ratio and also on the toroidal field reversal ratio, F.

The F-Θ universal curves depend also on the aspect ratio, especially in the sense that the largest deviation from the Taylor's curves are observed in the low aspect ratio RELAX device.

Finally, there are equilibrium reconstructions of the parallel current profile, although likely affected by relatively large error bars, that seem to show that the shape of this profile is relatively invariant with F.

In section II these results will be compared with the models predictions.

## II. *Models solutions at aspect ratio A=2 and comparison with RELAX data*

Before comparing the results of the models with the experimental data in RELAX it is useful to remind the main characteristics of the SHR and TR models, in particular as regard the free and adjustable parameters. More details can be found in [1].

### *II.1 Adjustable parameters in the models*

The SHR model has 3 free parameters: the aspect ratio A, the mode number n, and an adjustable exponent d [1].

The SHR model assumes a dominant helicity in the plasma and a minimization procedure for energy subjected to the conservation of total helicity (as in Taylor's theory) and also of an invariant related to the dominant mode expressed as :

$$K_1 = \frac{1}{2} \int_V \chi^d \mathbf{A} \cdot \mathbf{B} \, dV$$

where the integral is over the whole plasma volume, **A** and **B** are respectively the magnetic potential and the magnetic field and χ is the helical flux function of the mode defined as:

$$\chi = q_S \Psi - \Phi$$

where $q_s$ is the mode pitch, Ψ is the poloidal flux, while Φ is the toroidal flux, d is a positive integer. The new invariant in Eq. (1) corresponds therefore to the total helicity "weighted" over some power of the helical flux (and hence of the helicity) of the dominant mode. The underlying idea is that together with the total helicity, the invariant $K_1$ is also a well preserved quantity in a relaxation process dominated by a single mode. Note that the exponent d is needed in order to obtain solutions independent of the normalization of the magnetic field. This is clearly a very important and necessary physical constraint.

The behavior of the solutions with d was already analyzed in Ref. [1] and it was found that apparently the most appropriate choice to fit the data at A=4 was obtained by taking d=2.

An important point is that while within this theory, n, the mode number is free, the value of $\lambda$ (the parallel current) on axis, i.e. $\lambda(0)$, is an eigenvalue of the system. That is for each choice of (A, n, d) the RFP solutions only exist in some intervals of $\lambda(0)$ values. Once the correct $\lambda(0)$ is given the profiles for the magnetic field and the current density and therefore the entire $\lambda(r)$ profile are obtained. Each eigenvalue, $\lambda(0)$, corresponds to a solution with a different F and $\Theta$.

Beside the SHR model we have also the TR model that is obtained by assuming Bessel Functions solutions throughout the plasma region [1].

This model considers a flat $\lambda$ in the core region, between the axis and some matching radius, $r_c$, where the solution is exactly equal to the Taylor's one; beyond $r_c$ and till r=a i.e. the wall radius, the solution is given by a combination of Bessel Functions. At the matching radius it is possible to assume discontinuity (current sheet) or continuity of the solutions. To reduce the number of free parameters in [1] and also in the following, a continuous solution is assumed.

Therefore the free parameters of the TR model are the value of $\lambda(0)$ (that we name $\lambda_o$), the radius $r_c$ and a third parameter, $\lambda_1$, determining the solution in the interval $[r_c, a]$. Therefore also the TR model has 3 free parameters: ($\lambda_o, \lambda_1, r_c$).

In order to link the SHR model to the TR model and reduce, in this way, as much as possible the free parameters, it was assumed [1] that the value of $\lambda_o$ in the TR model is equal to the $\lambda(0)$ eigenvalues found in searching the SHR solutions. Furthermore it has been assumed that $\lambda_1 = 0.6 \lambda_o$ since with this choice reasonable profiles are obtained [1]. Therefore the only remaining free parameter in the TR model is $r_c$. For the case A=4 $r_c$ was set approximately 0.4 a, assuming, as observed in many experiments that the dominant modes are producing an island centered around this position, so that this could be roughly the position of a possible current sheet appearance. We will show later that this parameter has in fact a certain importance and should be chosen carefully especially for this low aspect ratio case.

## II.2 Comparison with the RELAX data

We start the comparison by looking at the F-Θ curves prediction, as shown in Fig.6.

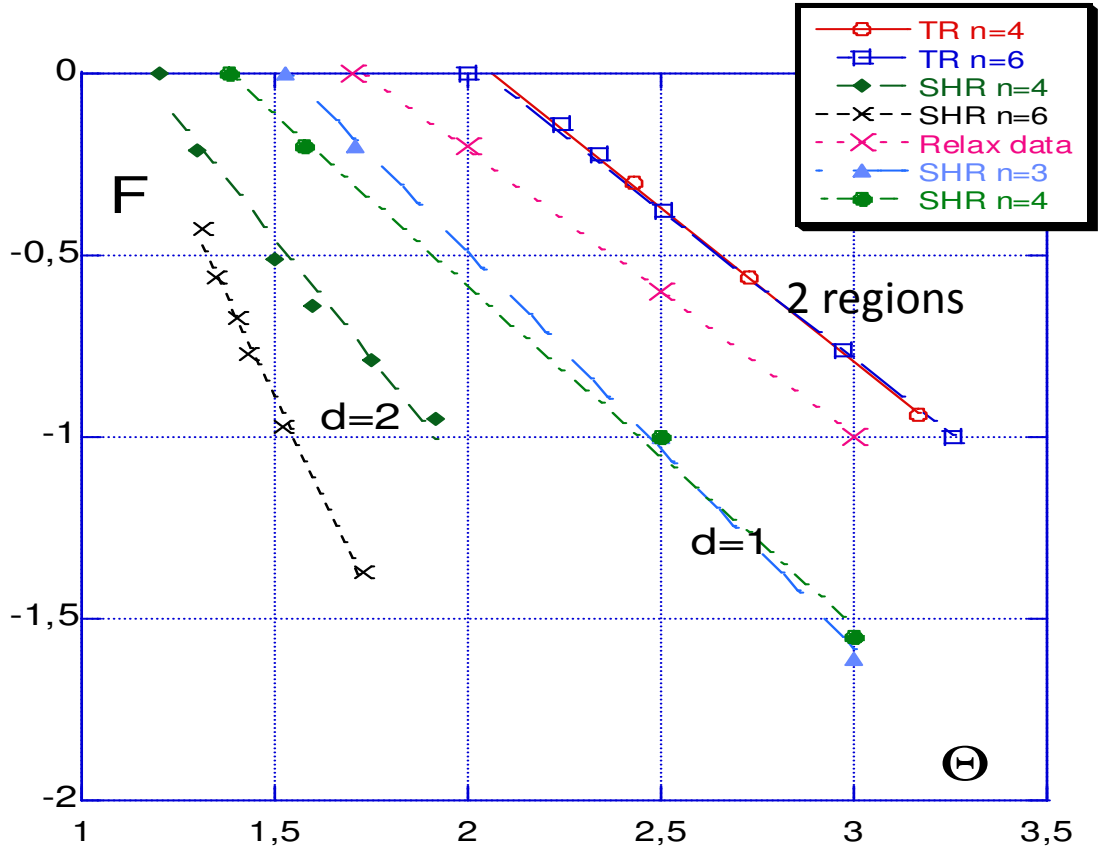

**Fig.6** F vs. Θ : experimental fit (crosses also labeled by exp); d=1 and d=2 SHR predictions with n=3 and n=4 and n=4 and n=6 respectively and the 2 region model predictions for n=4 and n=6.

It can be seen that for the SHR predictions with d=2 , differently from the medium aspect ratio case discussed in [1], the model curves are quite far from the experimental fit, with the n=6 shifting the curve to the left, while in the experiments all dominant modes are almost aligned along the same curve. This can also be seen in Fig.4(b) (section I) where the alignment in F and Θ for n=4 and n=6 dominant helicities is shown.

The situation seems a little better for the case with d=1, however for this case we were unable to find solutions with n higher than 4 at deep F values, again in contradiction with the experiment.

The best agreement, although not perfect, is in correspondence to the TR model. Note that the TR curve is obtained from the case with d=2 by assuming for the free parameters the quantitative link with the SHR model described in the previous paragraph.

Obviously beyond the comparison with averaged parameters, like F and $\Theta$, it is important also to look in more details the model predictions as regard in particular the $\lambda$ profile.

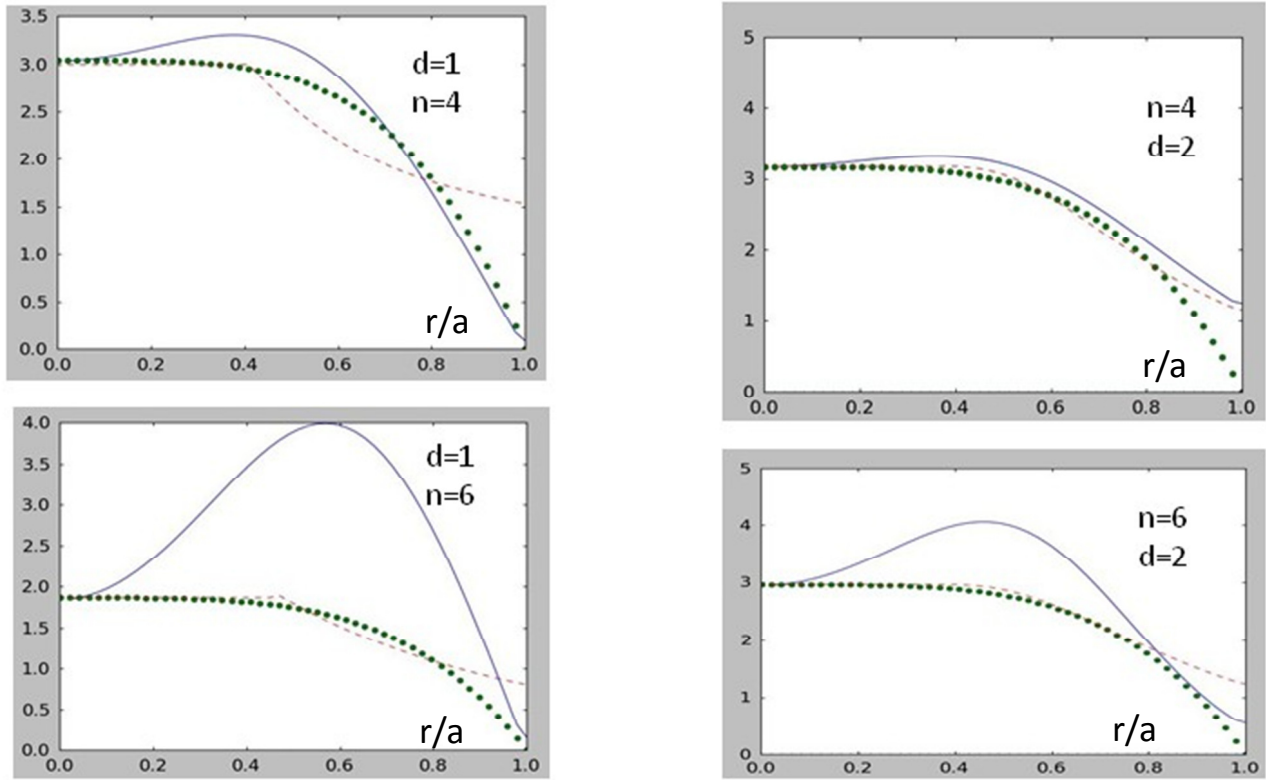

**Fig.7** $\lambda$ profiles vs normalized radius (r/a). Dotted lines correspond to experimental fit (see Fig.4); plain lines to SHR and dashed lines to the TR curves. Top curves refer to shallow and bottom curves to deeper F cases.

In Fig.7 the comparison is shown both for d=1 and d=2. It can be seen that there is a tendency, especially for the case with d=1, to develop a bump in the $\lambda$ profile by decreasing the F value. In fact for this case we were unable, as we already mention, to find solutions at very deep F for n=6.

It can be also seen, that apart a small region near the edge, when d is set to 2, the TR model is very close to the experimental fit (dotted lines), for both reversal ratios.

In order to take into account of the appearance of the bump in the parallel current profile at high F,

we modify the link between the TR and the SHR models described in section II.1. In particular instead of assuming $\lambda_o$ equal to the $\lambda(0)$ obtained by the SHR we set it equal to the average of the $\lambda$ profile in the region from the maximum to the axis, as shown in Fig.8.

In the shallow F cases, when $\lambda$ from SHR is flat in the core the two assumptions are equivalent. Instead when $\lambda$ obtained from the SHR model shows a maximum (as for deeper F's) the new assumption differs from the old one.

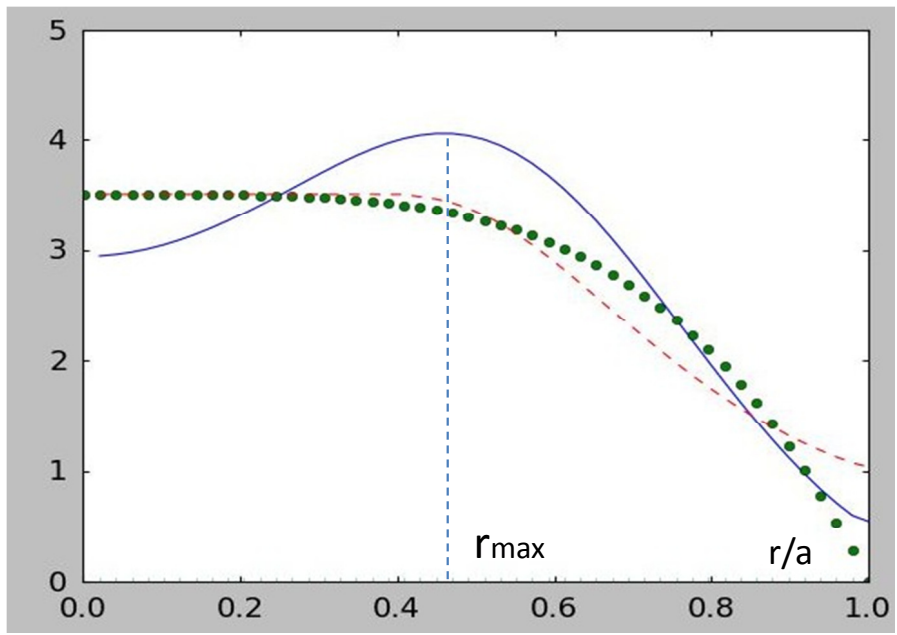

**Fig.8** $\lambda$ vs. normalized radius (r/a): the dotted line is the experimental fit, the plain line the SHR model (d=2) and the dashed line is the TR model. Note that the TR $\lambda(0)$ is now set to the average of the plain line (SHR model) between $r_{max}$ and r=0.

Another point that should be analyzed more carefully is the position of the matching radius in the TR model at low aspect ratio.

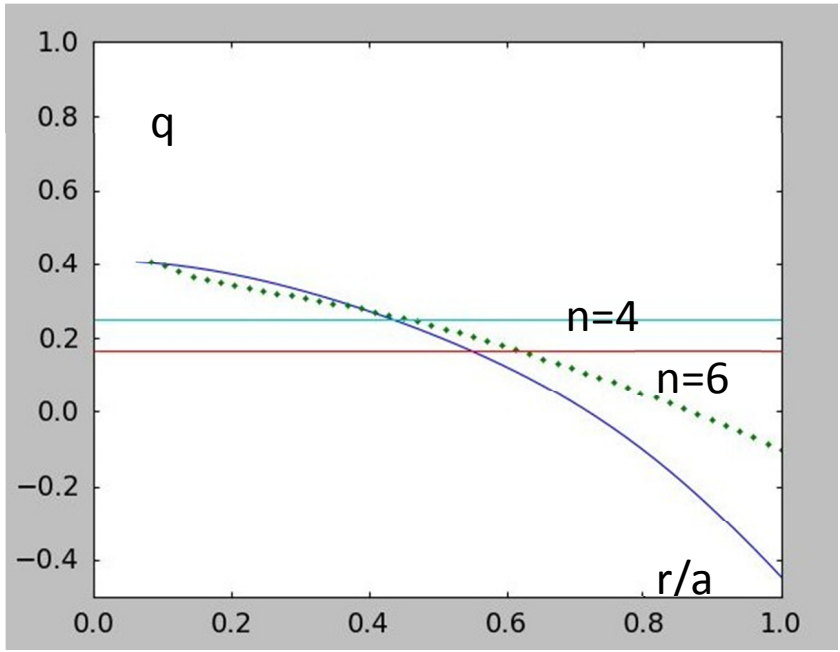

**Fig.9** q profile vs. normalized radius (r/a): plain line SHR model, crosses TR model. Resonances of n=4,6 modes are shown.

In Fig.9 the q profiles obtained with the SHR and TR models are shown. It can be seen that the resonances for the n=4,6 modes ( and therefore the possible position of a current sheet) are located at normalized radii between 0.5 and 0.6. Therefore the assumption about the position of the matching radius in the TR model should be revised with respect to the value $r_c$ =0.4 that was selected previously for A=4 [1]. By taking into account the results shown in Fig.9 we set the position of the matching radius at $r_c$ = 0.5.

Applying both these new assumptions we consider again the F-Θ curves, comparing the experimental fit with the curves obtained by the TR model (see Fig.10).

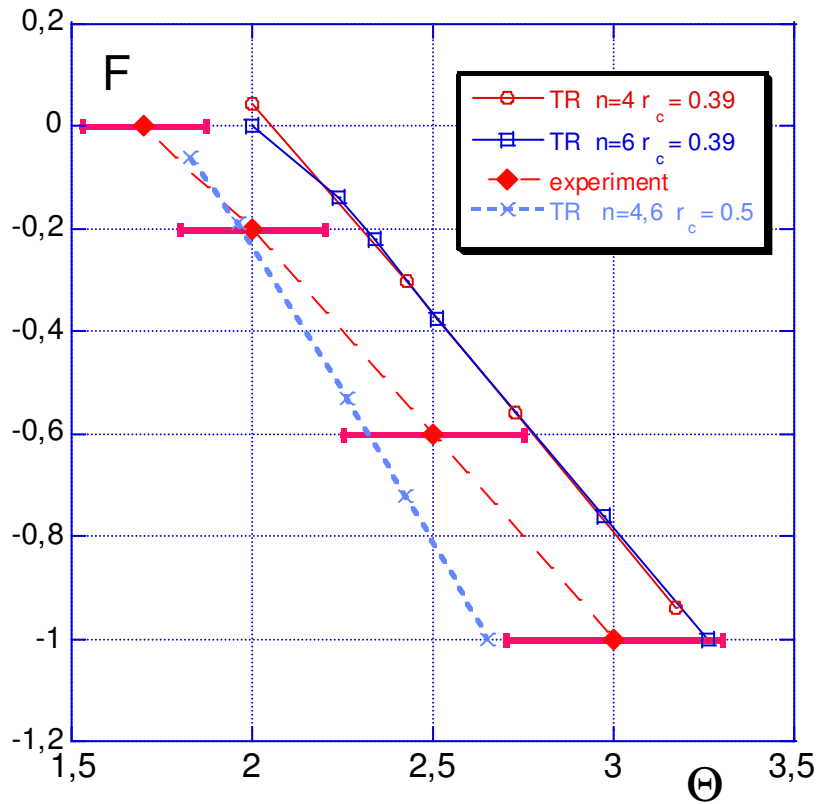

**Fig.10** F vs. Θ : experimental fit with error bars (red long dashed), TR model with $r_c$=0.4 (circles and squares), TR model with $r_c$=0.5 (light blue short dashed).

It can be seen that there is now a good agreement between the TR model and the experimental data within their typical error bars.

It is very important to remark that it is only by using together the SHR and the TR model that we could obtain a better fit of the data. The SHR eigenvalue $\lambda(0)$ (or as in this case the $\lambda$ average value in the core) and the position of the resonance of the dominant mode are the two essential ingredients used within the TR model.

The last point we want to address, is about the predictions that can be obtained from our models for the emergent dominant modes at the various F's.

Once again the separate use of the SHR and TR models could not give information about this important aspect, since in the SHR model the n value is a free parameter while the TR solution is built only after that the SHR state has been determined.

However in [1] it was proposed that by constructing a suitable measure of the "distance" between the two models, SHR and TR, in terms of the λ profile differences, this information could be extracted.

The following integral "error" in the parallel current profile was considered:

$$E(q_s) = \int (J_\parallel^{\lambda_{0,1}} - J_\parallel^{\chi})^2 \, dr$$

where the first term is the parallel current as deduced by the TR model (assuming all the links with the SHR model described above) and the second term is deduced directly from the SHR model. In [1] it was shown that the favorite/dominant n value at the different F's can be deduced by minimizing the above integral, which depends for each F from the selected n mode. The results are shown in Fig.11.

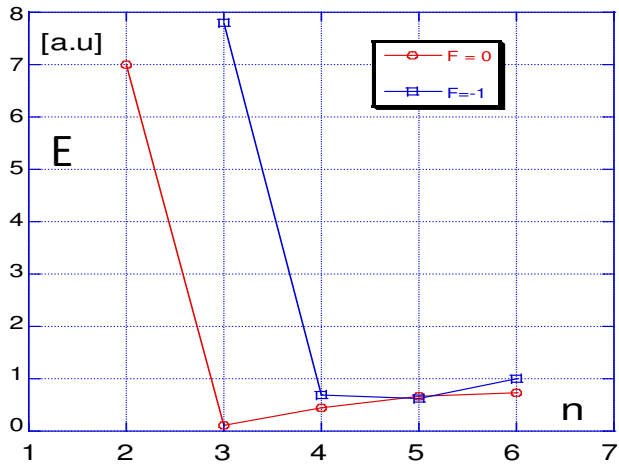

**Fig.11** Integral error E vs. n for shallow and deep F values.

It can be seen that at shallow F the minimum is obtained for n=3 and the next favorite mode is n=4, while for deep F, n=4,5,6 produce all very similar values of the error.

These predictions are in a reasonable agreement with the experimental results shown in Fig.4(a), where it can be seen that at shallow F the highest persistence is shared by n=3 and n=4 modes, with a dominance of the n=4, while at deep F the likelihood of n=5,6 is very similar. As discussed in section I the persistence of the dominant modes is also quite low at deep F in the experiment. We should also note that experimentally in RELAX the n=3 emerges mainly in slightly non-reversed cases (Fig.4(a)), although as discussed above the precise evaluation of F could be problematic near F=0, due to systematic toroidal asymmetries in the device.

## III. Perturbation data and Helical Grad-Shafranov solver

In a recent paper [19] a method has been proposed to find the solution of a Helical Grad-Shafranov (HGS) equation that can describe, starting from the axisymmetric solution obtained from the SHR model, the perturbed dominant helical mode.

In this section we compare the prediction of the model with some existing experimental data obtained in RELAX. In particular the aim is to calculate the components of the helical magnetic field and to compare them with data obtained by insertable magnetic probes. The data for the perturbation are obtained at shallow F (around 0) with a dominant n=4 mode. Since at low F and for n=4 the prediction of the SHR model with d=1 are matching not too bad the F-Θ data, as seen in the previous paragraph, we choose this value in the following.

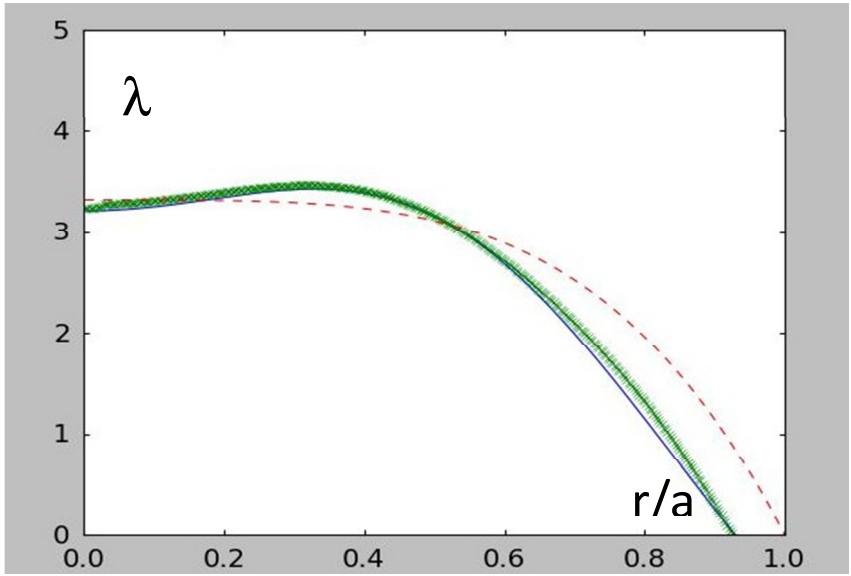

**Fig.12** $\lambda$ profiles vs normalized radius: contimous line is the SHR solution and crosses corresponds to the helical Grad-Shafranov modified solution. Dashed line is the reference fit to experimental data.

In Fig.12 the $\lambda$ profiles vs. radius from the SHR model, the Grad-Shafranov solver [19] and the experimental fit (the dashed line corresponds to the same fit of Fig.5 in section II) are shown for a case with F=-0.03 and Θ=1.61. Note that by letting $\lambda(a)$ (as done in[19]) not to be zero (see Fig.12) the fit of the experimental data in terms of F and Θ becomes excellent at shallow F and with n=4 (while still unsatisfactory for n=6 at deep F).

From the HGS as described in [19] the radial profile of the dominant mode magnetic field components can be calculated and compared with the data of the insertable probes, as shown in Fig.13. The experimental profiles were obtained as follows. Signals from the radial array of magnetic probes, which measured three components at 10 radial locations, were time-integrated, and a numerical band-pass filter (2 kHz < $f$ < 50 kHz) was used to extract only the fluctuating components. The ordinate is the fluctuating magnetic field amplitude normalized to the edge poloidal field [23].

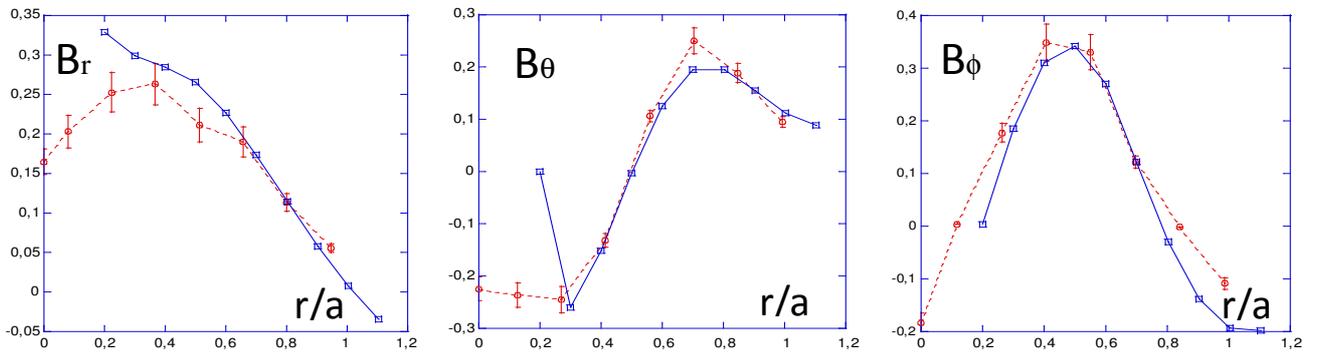

**Fig.13** From the left experimental radial, poloidal and toroidal perturbed magnetic field components (in red with error bars) compared to the n=4 (squares) dominant mode calculated from the HGS solver.

It should also be noted that the theoretical curves are multiplied by an arbitrary normalization factor and shifted radially of 0.2 of the normalized radius, to take into account the relatively big toroidal Shafranov shift present in the experiment (5 over 25 cm). It can be seen that the comparison is quite satisfactory. From the knowledge of the radial magnetic field perturbation it is also possible to reconstruct the island structure as shown in Fig.14.

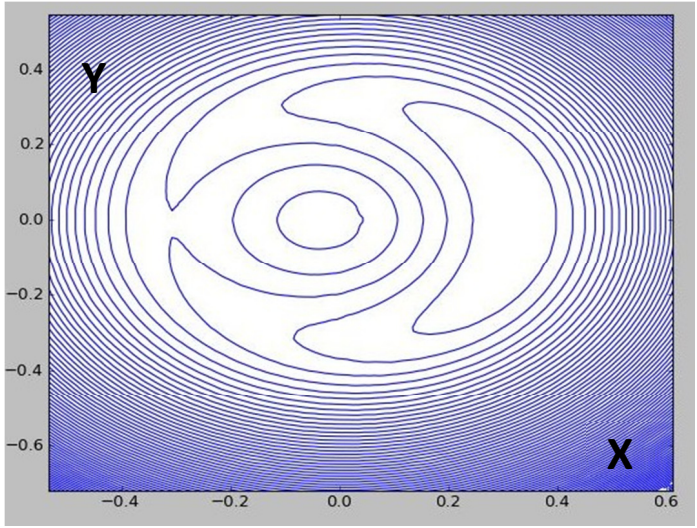

**Fig.14** Island structure of the n=4 dominant mode in the (X,Y) poloidal plane

This structure (shown here for the cylinder) should in principle also be moved outward according to the Shafranov shift to be compared with the experimental data.

**Discussion and Conclusions**

In this paper we have compared the outcome of the model proposed in [1] and in [19] to the low aspect ratio RELAX device data.

In this experiment SH helicity states are observed, as in all RFPs, both at shallow and deep reversal. However, as discussed in section I, RELAX is a small machine with generally quite fast time scales, therefore the emergence of dominant modes is only observed transiently and for relatively short time windows. This is especially true for the deep reversed plasmas and the highest n values.

Therefore the SH states in RELAX are far to be stationary and the comparison with the theoretical models should be taken with some caution, since for example in the SHR model the plasma configurations correspond, by construction, to stable equilibrium states.

Nevertheless we compared the predictions of the SHR and TR models (see previous sections) in terms of the F-Θ universal curves, first proposed within the Taylor's relaxation theory [11], describing in a synthetic way the RFP operational states.

We find that the SHR model predictions are not good enough, at least over the whole range of F and Θ, differently from what we obtained at higher aspect ratio (A=4) in Refs.[1,19]. One exception is represented by the shallow reversal data that are satisfactorily reproduced also by the SHR model with n=3 or 4 and d=1, particularly when λ(a) (the parallel edge current) is not forced to be exactly zero, as done previously [1].

We should also mention that the experimental SH states in positive F region may be different in nature from those in reversed field regions; in the former case the SH states (namely the n=3 or sometimes the n=4 modes) appear more during "transient" phases; that is, they tend to appear in the F-increasing phase toward more positive direction (F>0 and dF/dt>0). In particular, it should be kept in mind that in the positive F region, a non-resonant kink mode could in principle become unstable [25], which is clearly not the subject of this paper.

However, since n=3 (and n=4) states at shallow reversal are predicted also by the SHR model, it may also be that the experimental measurements near F=0 could be affected by errors due to toroidal asymmetries, which may produce a shift on the F estimated values.

The TR model seems to fit better the experimental data over the whole F-Θ range. The fit improves by taking into account into the free parameters of the TR model of the appearance in the SHR model of a bump in the parallel current profile, λ, and also of the fact that the resonances of the n=4,6 modes moves at larger radii for the solutions obtained with A=2. In particular these results suggest, on one hand, to consider in the TR model the average of the current in the core (between the magnetic axis and the current maximum) as obtained from the SHR solution and, on the other hand, to increase the matching radius, $r_c$, of the TR solution, outwards with respect to the previous intermediate aspect ratio cases (A= 4 ). With these changes, the fit of the experimental data by the TR model becomes very reasonable.

Beside the F-Θ curves we compare the models predictions for the λ profiles and again the TR model fits very well the data, although at shallow reversal and for low n's also the SHR model could be considered good enough. However in the RELAX experiment there is no sign of a bump in the parallel current at deep F as predicted by the SHR relaxation model, although this point should be better addressed in the context of the sensitivity of the experimental equilibrium reconstruction techiniques. As shown here the bump is much more pronounced by taking d=1 in the SHR model at deep F. However as noted in [19] the most general solution of the SHR model is a sum over different d's . Therefore it could be possible that to better match the data it would be necessary to let the free exponent d to vary from 1 to 2 going from shallow to deep F values. As discussed in [19] allowing this possibility will make very difficult the simultaneous numerical search of the corresponding three eigenvalues of the SHR model. Therefore we limit the search for cases that have either d=1 or d=2 over the whole parameter space, as already done elsewhere [1,19].

We check, further, in this paper, that the dominant modes at various F's are consistently predicted by the theory applying the same integral error minimization techniques proposed in [1]. This approach seems to work well also for the present case. At low aspect ratio however these minima are shown not to be a strong varying function with n, as somehow suggested also by the experimental results.

The fact that the TR model predicts more accurately the experimental trends and, in particular that, the experimental λ profiles do not show the maxima obtained within the SHR model, could be due to a residual stochasticity in the core of the RELAX plasma that tends to flatten the parallel current there. This could be, on the other hand, coeherent with the relatively fast time scales observed in the experiment. As discussed already, the RELAX plasmas are far to show stationary SH phases, therefore it is maybe not surprising that the data do not match completely with a theory that is consistent with a steady state plasma with well conserved invariants. This caveat becomes more and more important, as shown in section I, for high n modes at deep reversal, where the experimental SH states are lost very quickly. What can be presently observed in the RELAX experiment at deep F values are more like trends and indications of what would be the preferred/likely mode number in the case in which a well developed quasi-stationary SH state would take place.

A more radical point of view about the failures of the SHR cylindrical theory, could be that it is not able to fully reproduce a case at A=2, where toroidal effects could play an important role.

Preliminary results, by adding a first order toroidal corrections to the SHR solutions, show that the F-Θ curves do not significantly change. On the other hand , as it was shown in section III, the prediction of a cylindrical helical Grad-Shafranov solver (initialized by the SHR solution) matches very well the measured perturbed magnetic fields (once that the experimental horizontal Shafranov shift is taken into account) in RELAX. In any case future studies using toroidal codes are already planned to address more extensively this issue.

The reconstruction of the perturbed helical magnetic fields would also, in principle, allow a detailed comparison with experimental tomographic data, whenever available, especially for what concerns the radial position of the island structure associated with the dominant mode. This topic is left for future investigations.

**References:**

**[1]** R. Paccagnella, Phys. of Plasmas **23** (2016) 092512.

**[2]** S. Masamune *et al,* J. Phys. Soc. Japan 76 ( 2007) 123501.

**[3]** Cappello S., Paccagnella R., Phys. Fluids **B 4** (1992) 611.

**[4]** Finn J.M., Nebel R.A., Bathke C., Phys. Fluids **B 4** (1992) 1262.

**[5]** Hirano Y. ,Yagi Y., Maejima Y., Shimada T. , Hirota I., Plasma Phys. Control. Fusion **39** (1997) A393.

**[6]** Martin P**.,** Plasma Phys. Control. Fusion **41** (1999) A247.

**[7]** L. Marrelli, P. Martin, G. Spizzo, P. Franz, B. E. Chapman, D. Craig, J. S. Sarff, T. M. Biewer, S. C. Prager, and J. C. Reardon, Phys. Plasmas **9** (2002) 2868.

**[8]** P. Piovesan, G. Spizzo, Y. Yagi, H. Koguchi, T. Shimada, Y. Hirano, and P. Martin, Phys. Plasmas **11** (2004) 151.

**[9]** P. Franz, L. Marrelli, P. Piovesan, I. Predebon, F. Bonomo, L. Frassinetti, P. Martin, G. Spizzo, B. E. Chapman, D. Craig, and J. S. Sarff, Phys. Plasmas **13** (2006) 012510.

**[10]** Ortolani S. and the RFX team., Plasma Phys. Control. Fusion **48** (2006) B371.

**[11]** J. B. Taylor, Phys. Rev. Lett**. 33** (1974) 1139.

**[12]** J.B. Taylor, Rev. Mod. Phys. **58** (1986) 751.

**[13]** R. Paccagnella , Nucl. Fusion **56** (2016) 046010.

**[14]** P. R. Brunsell, D. Yadikin, D. Gregoratto, R. Paccagnella, et.al. Phys. Rev. Lett. **93** (2004) 225001.

**[15]** R. Paccagnella, S. Ortolani, P. Zanca, A. Alfier, T. Bolzonella, L. Marrelli, M. E. Puiatti, G. Serianni,D. Terranova, M. Valisa et al. Phys. Rev. Lett. **97 (**2006) 075001.

**[16]** A. Bhattacharjee, R.L. Dewar, D.A. Monticello, Phys. Rev. Lett. **45** (1980) 347.

**[17]** A. Bhattacharjee, R.L. Dewar, Phys. Fluids **25** (1982) 887.

**[18]** A. Bhattacharjee, R.L. Dewar, A.H. Glasser, M.S. Chance, J.C. Wiley, Phys. Fluids **26** (1983) 526.

**[19]** R. Paccagnella, Phys. of Plasmas **25** (2018) 022112.

**[20]** K. Oki, A. Sanpei, H. Himura, S. Masamune, Trans. Fus. Sci. Technol. **63** (2013) 386.

**[21]** D. F. Escande, P. Martin, S. Ortolani, A. Buffa, et. al. Phys. Rev. Lett. **85** (2000) 1662.

**[22]** L. Frassinetti, P.R. Brunsell, J.R. Drake, S. Menmuir and M. Cecconello, Phys. of Plasmas **14** (2007) 112510.

**[23]** K. Oki, D. Fukahori, K. Deguchi, S. Nakaki, A. Sanpei. H. Himura, S. Masamune, R. Paccagnella, Plasma Fusion Res. 7, 1402028 (2012).

**[24**] R. Ikezoe, K. Oki, T. Onchi, Y. Konishi, M. Sugihara, S. Fujita, A. Sanpei, H. Himura , S . Masamune, Plasma Phys. Control. Fusion **53** (2011) 025003.

**[25]** Y. Hirano, R. Paccagnella, H. Koguchi, L. Frassinetti, H. Sakakita, S. Kiyama, and Y. Yagi, Phys. of Plasmas **1 (**2005) 112501.